\documentclass[twocolumn,prb,showpacs]{revtex4}

\usepackage{graphicx}
\usepackage{dcolumn}
\usepackage{bm}
\usepackage{amsmath}

\begin{document}

\title{Anisotropic magnetic and superconducting properties of
aligned weak-ferromagnetic superconductor RuSr$_2$RCu$_2$O$_8$ (R =
rare earths)}
\author{B. C. Chang}
\author{C. H. Hsu}
\author{M. F. Tai}
\author{H. C. Ku}%
 \email{hcku@phys.nthu.edu.tw}
 \affiliation{Department of Physics, National Tsing Hua University,
Hsinchu 30013, Taiwan, Republic of China}
\author{Y. Y. Hsu}
\affiliation{Department of Physics, National Taiwan Normal
University, Taipei 11677, Taiwan, Republic of China}

\date{\today}



\begin{abstract}

The powder alignment method is used to investigate the anisotropic
physical properties of the weak-ferromagnetic superconductor system
RuSr$_2$RCu$_2$O$_8$ (R = Pr, Nd, Sm, Eu, Gd, Gd$_{0.5}$Dy$_{0.5}$).
The RuSr$_2$GdCu$_2$O$_8$ Ru-1212 cuprate is a weak-ferromagnetic
superconductor with a magnetic ordering of Ru moments at T$_N$(Ru) =
131 K, a superconducting transition in the CuO$_2$ layers at T$_c$ =
56 K, and a low temperature Gd antiferromagnetic ordering at
T$_N$(Gd) = 2.5 K. Due to weak magnetic anisotropy of this
tetragonal system, highly $\it{c}$-axis aligned microcrystalline
powder (diameter $\sim$ 1-10 $\mu$m) in epoxy can be obtained only
for R = Eu and Gd through the field-rotation powder alignment method
where $\it{c}$-axis is perpendicular to the aligned magnetic field
B$_{a}$ = 0.9 T and parallel to the rotation axis. For smaller rare
earth compound R = Gd$_{0.5}$Dy$_{0.5}$, powder alignment can be
achieved using the simple field powder alignment method where
$\it{c}$-axis is partially aligned along the aligned magnetic field.
No powder alignment can be achieved for larger rare earths R = Pr,
Nd or Sm due to the lack of magnetic anisotropy in these compounds.
The anisotropic temperature dependence of magnetic susceptibility
for the $\it{c}$-axis aligned powders exhibit weak anisotropy with
$\chi_{c} > \chi_{ab}$ at room temperature due to anisotropic rare
earth, Eu and Gd, contribution and crossover to $\chi_{c} <
\chi_{ab}$ below 190 K where strong Ru anisotropic short-range
exchange interaction overtakes the rare earth contribution.
Anisotropic diamagnetic superconducting intragrain shielding signal
of aligned microcrystalline RuSr$_2$GdCu$_2$O$_8$ powder-in-epoxy
below vortex lattice melting temperature at 39 K in 1-G field is
much weaker than the intergrain polycrystalline bulk sample signal
due to the small grain size (d $\sim$ 1-10 $\mu$m), long penetration
depth ($\lambda_{ab} \sim$ 0.55 $\mu$m, $\lambda_{c} \sim$ 0.66
$\mu$m) and the two-dimensional (2D) character of CuO$_2$ layers.

\end{abstract}

\maketitle

\section{Introduction}

Magnetic superconductivity has attracted much research attention
since it was reported in the strongly-correlated
RuSr$_{2}$RCu$_{2}$O$_{8}$ Ru-1212 cuprate system (R = Sm, Eu, Gd,
Y) with the tetragonal space group P4/mbm. \cite{p1,p2,p3,p4,p5} The
Ru magnetic moments order weak-ferromagnetically (WFM) with ordering
temperature T$_{N}(Ru) \sim$ 130 K. High-T$_{c}$ superconductivity
occurs in the quasi-2D CuO$_{2}$ bi-layers from doped holes with
maximum superconducting transition onset T$_{c}$(onset) $\sim$ 60 K
for R = Gd and coexists with the WFM order. A crossover from
anisotropic 2D-like to less-anisotropic 3D-like structure was
observed near R = Sm, along with a metal-insulator transition. No
superconductivity can be detected for the Mott insulators R = Pr and
Nd.

Since the oxygen content for all samples is close to eight after
oxygen annealing, the variation of T$_{c}$ with rare-earth ions
indicates a self-doping of electrons from CuO$_{2}$ layers to
RuO$_{6}$ layers. Such self-doping creates hole carriers in
CuO$_{2}$ layers and conduction electrons in RuO$_{6}$ layers. The
Ru L$_{3}$-edge X-ray absorption near-edge spectrum (XANES) of
RuSr$_{2}$GdCu$_{2}$O$_{8}$ indicates that Ru valence is basically
at Ru$^{5+}$ (4d-t$_{2g}^{3}$, S = 3/2) state with small amount
($\sim$20 $\%$) of Ru$^{4+}$ (4d-t$_{2g}^{4}$, S = 1 in low spin
state) which corresponds to doped electrons. \cite{p6} The strong
antiferromagnetic superexchange interaction between Ru$^{5+}$
moments is responsible for the basic G-type antiferromagnetic order
observed in the neutron diffraction study. \cite{p7} The weak
ferromagnetic component observed from magnetic susceptibility and
NMR spectrum is probably due to weak-ferromagnetic double-exchange
interaction through doped conduction electrons in the metallic
RuO$_{6}$ layers.

Since the magnetic superexchange and double-exchange interaction is
anisotropic in general, the study of anisotropic physical properties
is crucial for this quasi-2D system. In this report, we align the
microcrystalline RuSr$_{2}$RCu$_{2}$O$_{8}$ (R = rare earths) powder
($\sim$1-10 $\mu$m) in magnetic field to investigate the anisotropic
properties.

\section{experimental}

The stoichiometric RuSr$_{2}$GdCu$_{2}$O$_{8}$ bulk sample was
synthesized by the standard solid-state reactions. High-purity
RuO$_{2}$ (99.99 $\%$), SrCO$_{3}$ (99.9 $\%$), Gd$_{2}$O$_{3}$
(99.99 $\%$) and CuO (99.99 $\%$) preheated powders with the nominal
composition ratio of Ru:Sr:Gd:Cu = 1:2:1:2 were well mixed and
calcined at 960$^{\circ}$C in air for 16 hours. The calcined powders
were then pressed into pellets and sintered in flowing N$_{2}$ gas
at 1015$^{\circ}$C for 10 hours to form Sr$_{2}$GdRuO$_{6}$ and
Cu$_{2}$O precursors. The sintered pellets were then heated at
1060-1065$^{\circ}$C in flowing O$_{2}$ gas for 7 days to form the
Ru-1212 phase and slowly furnace cooled to room temperature with a
rate of 15$^{\circ}$C per hour. For powder alignment in magnetic
field, samples were ground into powders with an average
microcrystalline grain size of 1-10 $\mu$m and mixed with epoxy
(4-hour curing time) in a quartz tube ($\phi$ = 8 mm) with the ratio
of powder:epoxy = 1:5 then immediately put into the alignment field
environments (simple field or rotation-field alignment).

\section{Results and Discussion}

For simple powder alignment, the mixture was placed in a 14-T
superconducting magnet at room temperature in flowing N$_{2}$ gas
and slowly hardened overnight as shown in figure 1. The powder X-ray
diffraction pattern of three typical aligned powder-in-epoxy samples
RuSr$_{2}$RCu$_{2}$O$_{8}$ (R = Sm, Eu, Gd$_{0.5}$Dy$_{0.5}$) are
shown collectively in figure 2. For R = Sm (as well as for R = Pr
and Nd), no magnetic alignment can be achieved. The lack of magnetic
anisotropy may closely relate to the variation of tetragonal lattice
parameters where $\it{c}$/3 $\sim \it{a}$/$\sqrt{2}$ for R = Sm with
$\it{a}$ = 0.5448 nm and $\it{c}$ = 1.1560 nm (space group P4/mbm)
as shown in figure 3. For R = Eu (as well as for R = Gd), partial
($\sim$ 90$\%$) $\it{ab}$-plane aligned along alignment magnetic
field B$_{a}$ is observed through the appearance of enhanced
($\it{hk}$0) diffraction lines. A small amount of SrRuO$_{3}$
impurity is presented. The $\it{ab}$-plane alignment may be due to
the fact that $\it{c}$/3 $> \it{a}$/$\sqrt{2}$ for R = Eu ($\it{a}$
= 0.5435 nm, $\it{c}$ = 1.1572 nm). For metastable compound R =
Gd$_{0.5}$Dy$_{0.5}$ near the phase boundary with some unreacted
precursor Sr$_{2}$RRuO$_{6}$, partially $\it{c}$-axis alignment
along B$_{a}$ is detected with enhanced (00$\it{l}$) lines due to
$\it{c}$/3 $< \it{a}$/$\sqrt{2}$ in this compound ($\it{a}$ = 0.5426
nm, $\it{c}$ = 1.1508 nm).

\begin{figure}
\includegraphics{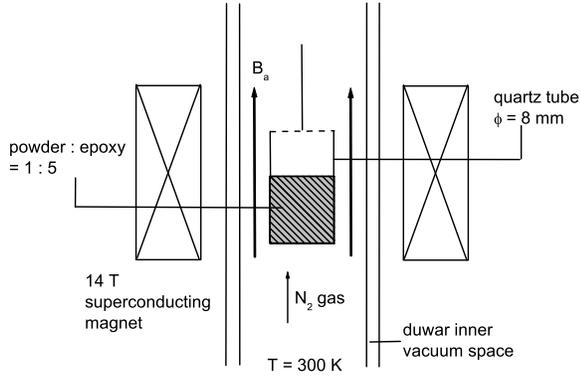}
\caption{\label{label} Schematic diagram for the magnetic field
powder alignment method in a 14 T superconducting magnet at 300 K.}
\end{figure}

\begin{figure}
\includegraphics{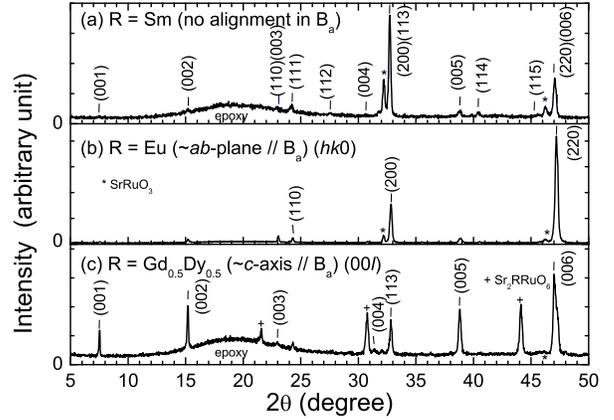}
\caption{\label{label} Powder X-ray diffraction patterns for
RuSr$_{2}$RCu$_{2}$O$_{8}$ aligned powder. (a) R = Sm, (b) R = Eu,
(c) R = Gd$_{0.5}$Dy$_{0.5}$.}
\end{figure}

\begin{figure}
\includegraphics{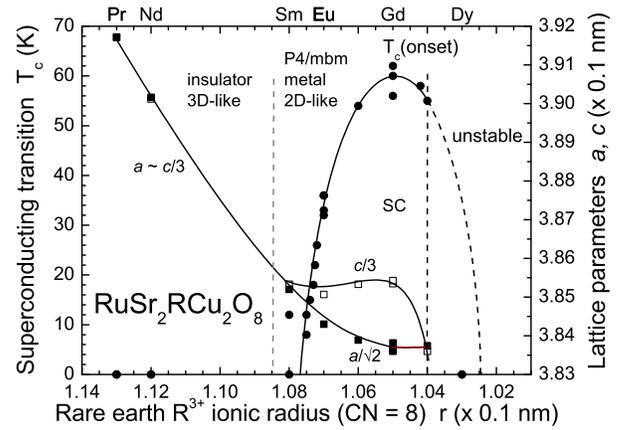}
\caption{\label{label} The variation of superconducting transition
T$_{c}$ and tetragonal lattice parameters $\it{a}$, $\it{c}$ with
rare earth ionic radius R$^{3+}$ for RuSr$_{2}$RCu$_{2}$O$_{8}$
system (R = Pr-Dy).}
\end{figure}

The phase diagram in figure 3 indicates a structural crossover from
less-anisotropic 3D-like ($\it{c}$/3 $\sim \it{a}$) to anisotropic
2D-like structure ($\it{c}$/3 $\ne$ $\it{a}$/$\sqrt{2}$) near R =
Sm, along with an insulator-to-metal transition. Superconductivity
appears only in the quasi-2D metallic region with resistivity onset
transition temperature T$_{c} \sim$ 0 for R = Sm, T$_{c}$ = 36 K for
R = Eu, T$_{c}$ = 56 K for Gd, and T$_{c}$ = 55 K for metastable R =
Gd$_{0.5}$Dy$_{0.5}$.

For R = Eu with $\it{ab}$-plane aligned along B$_{a}$, $\it{c}$-axis
can be in any direction within the plane perpendicular to B$_{a}$.
To obtain the $\it{c}$-axis aligned powder, a field-rotation
alignment method is used as shown in figure 4. Since $\it{ab}$-plane
is fixed along B$_{a}$, the rotation of quartz tube (10 rpm)
perpendicular to B$_{a}$ forces the microcrystalline $\it{c}$-axis
to have no choice but to be aligned along the rotation axis. The
powder X-ray diffraction patterns of RuSr$_{2}$EuCu$_{2}$O$_{8}$
random powder ($\it{hkl}$), partially $\it{ab}$-plane aligned along
B$_{a}$ ($\it{hk}$0), and highly $\it{c}$-axis aligned along the
rotation axis (00$\it{l}$) are shown collectively in figure 5. The
relative intensity of the enhanced (00$\it{l}$) lines and (113) line
indicates a $\sim$ 90$\%$ c-axis alignment in this field-rotation
aligned powder-in-epoxy sample.

\begin{figure}
\includegraphics[scale=1]{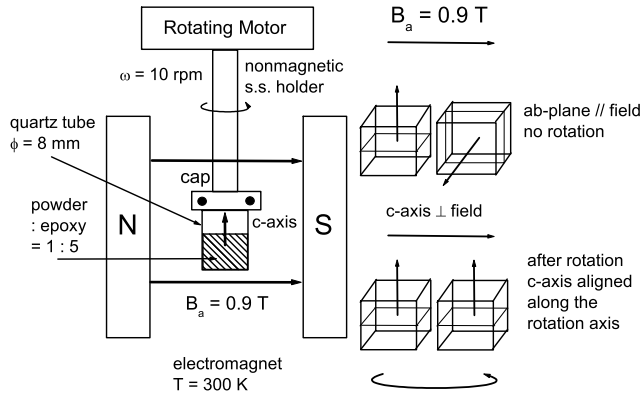}
\caption{\label{label} Schematic diagram for the field-rotation
powder alignment method with $\it{c}$-axis perpendicular to aligned
magnetic field and along the rotation axis.}
\end{figure}

\begin{figure}
\includegraphics{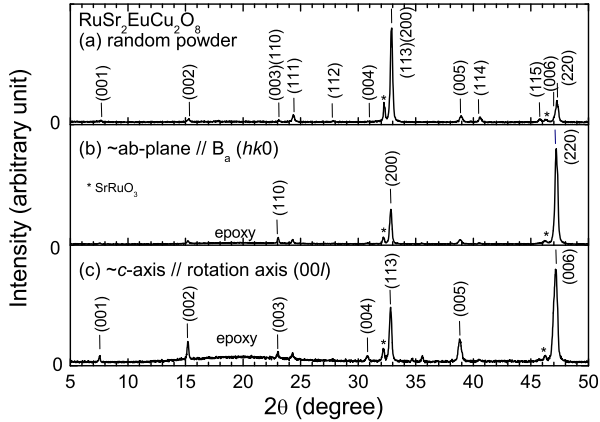}
\caption{\label{label} powder X-ray diffraction patterns for
RuSr$_{2}$EuCu$_{2}$O$_{8}$. (a) random powder, (b) $\it{ab}$-plane
aligned along B$_{a}$, and (c) $\it{c}$-axis aligned along the
rotation axis.}
\end{figure}

Figure 6 shows the field dependence of paramagnetic moment of
RuSr$_{2}$GdCu$_{2}$O$_{8}$ aligned powder up to 7 T at 300 K. Since
$\it{ab}$-plane is aligned along the magnetic field in the alignment
procedure at room temperature, magnetic anisotropy of $\chi_{ab}
> \chi_{c}$ at 300 K is expected. However, 300 K m-B$_a$ data showed weak magnetic anisotropy with
linear paramagnetic magnetic moment m$_{ab} \sim$ 0.95 m$_{c}$ or
susceptibility $\chi$ = m/B$_{a}$ with $\chi_{ab} < \chi_{c}$.

The anisotropic temperature dependence of logarithmic molar magnetic
susceptibility of RuSr$_2$GdCu$_2$O$_8$ c-axis aligned powder in 1-T
applied magnetic field is shown in figure 7. A crossover from
$\chi_{ab} < \chi_{c}$ at 300 K to $\chi_{ab} > \chi_{c}$ at lower
temperature was observed around 185 K, followed by a
weak-ferromagnetic ordering at T$_{N}$(Ru) = 131 K.

\begin{figure}
\includegraphics{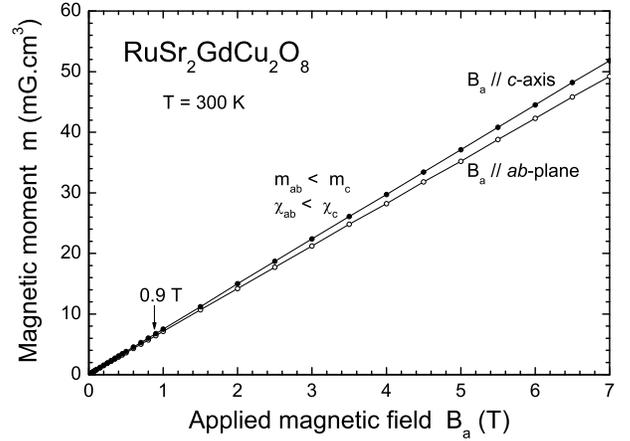}
\caption{\label{label} The field dependence of paramagnetic moment
of RuSr$_{2}$GdCu$_{2}$O$_{8}$ aligned powder up to 7 T at 300 K.
Linear paramagnetic magnetic moment.}
\end{figure}

\begin{figure}
\includegraphics{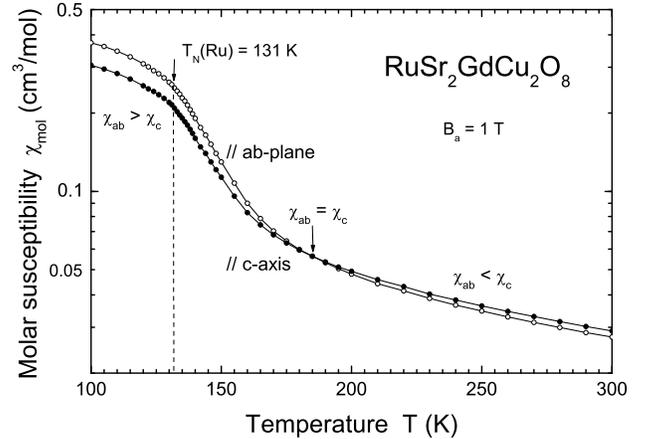}
\caption{\label{label} Temperature-dependence of logarithmic molar
magnetic susceptibility $\chi_{ab}$ and $\chi_{c}$.}
\end{figure}

The magnetic anisotropy of $\chi_{ab} < \chi_{c}$ observed at 300 K
is mainly due to the contribution of magnetic Gd$^{3+}$ ions (J =
7/2). The anisotropy of $\chi_{ab}$(Gd) $<$ $\chi_{c}$(Gd) is from
the tetragonal GdO$_{8}$ cage with anisotropic $\it{g}$-factor
$\it{g}_{ab} < \it{g}_{c}$, but with little 4$\it{f}$ wavefunction
overlap with the neighbor oxygen 2$\it{p}$ orbital.

Although there are three types of magnetic moments in this magnetic
superconductor: Ru$^{5+}$ (S = 3/2) with doped electrons or
Ru$^{4+}$ (S = 1), Cu$^{2+}$ (S = 1/2) with doped holes, and
Gd$^{3+}$ moment (J = 7/2), not all moments have the same
contribution in powder alignment. In the aligned magnetic field,
anisotropic orbital wavefunction is tied to the spin direction, and
a strong spin-orbital related short-range anisotropic exchange
interaction at 300 K should dominate the magnetic alignment. In the
present case, it is believed that Ru moment with the strong
short-range anisotropic double-exchange/superexchange interaction
along the $\it{ab}$-plane due to the Jahn-Teller distortion of
RuO$_{6}$ octahedron with $\chi_{ab}$(Ru) $>$ $\chi_{c}$(Ru) is the
dominant factor for $\it{ab}$-plane alignment along B$_{a}$ at 300
K. The shorter Ru-O(1) bond length in the tetragonal $\it{ab}$-basal
plane provides strong 4d$_{xy}$(Ru)-2p$_{x/y}$(O(1))-4d$_{xy}$(Ru)
wavefunction overlap. This exchange interaction increases with
decreasing temperature, and eventually total $\chi_{ab} > \chi_{c}$
was observed below 185 K as expected. The weak-ferromagnetic state
below 131 K is due to the long range order of this anisotropic
double-exchange/superexchange interaction.

\begin{figure}
\includegraphics{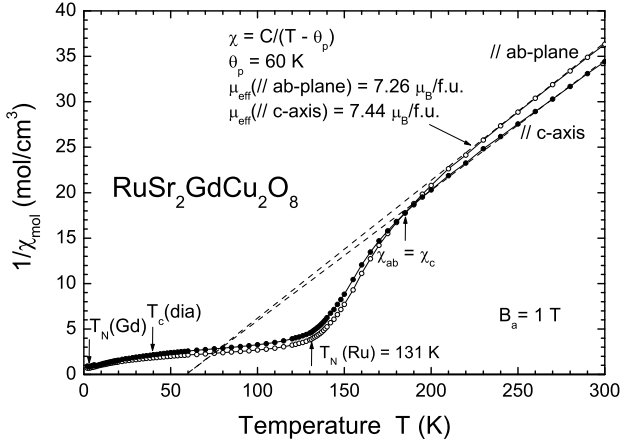}
\caption{\label{label} Reciprocal molar magnetic susceptibility
1/$\chi_{ab}$ and 1/$\chi_{c}$ for aligned
RuSr$_{2}$GdCu$_{2}$O$_{8}$ powder.}
\end{figure}

The reciprocal molar magnetic susceptibility 1/$\chi_{ab}$ and
1/$\chi_{c}$ of RuSr$_{2}$GdCu$_{2}$O$_{8}$ aligned powder are shown
in figure 8. A Curie-Weiss behavior $\chi$= C/(T - $\theta_{p}$) was
observed in the high temperature paramagnetic region above 200 K
with a Curie-Weiss intercept $\theta_{p}$ = 60 K for both field
orientations and the anisotropic effective magnetic moment
$\mu_{eff}^{c}$ = 7.44 $\mu_{B}$ per formula unit along the
$\it{c}$-axis, and $\mu_{eff}^{ab}$ = 7.26 $\mu_{B}$ per formula
unit along the $\it{ab}$-plane. Diamagnetic superconducting
transition T$_{c}$(dia) at 39 K and Gd ordering temperature
T$_{N}$(Gd) = 2.5 K are not clearly seen in this high applied field.

\begin{figure}
\includegraphics{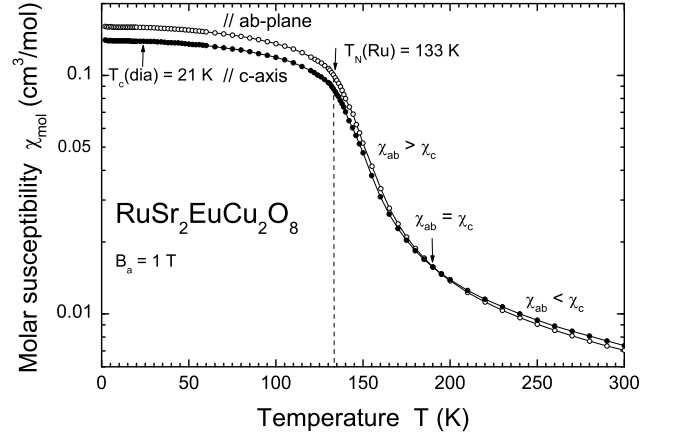}
\caption{\label{label} The temperature-dependence of logarithmic
molar magnetic susceptibility $\chi_{ab}$ and $\chi_{c}$ of aligned
RuSr$_{2}$EuCu$_{2}$O$_{8}$ powder in 1 T applied magnetic field.}
\end{figure}

The temperature dependence of logarithmic molar magnetic
susceptibility of aligned RuSr$_2$EuCu$_2$O$_8$ powder along
$\it{c}$-axis and $\it{ab}$-plane in 1-T applied magnetic field are
shown in figure 9. Similar to R = Gd compound, weak paramagnetic
anisotropy of $\chi_{ab}$ = 0.95 $\chi_{c}$ was observed at 300 K,
and a crossover to $\chi_{ab}> \chi_{c}$ was detected below 190 K
with a weak-ferromagnetic ordering temperature T$_{N}$(Ru) = 133 K.
Superconducting diamagnetic signal T$_{c}$(dia) at 21 K (resistivity
zero point) is very weak in large 1-T applied field.

\begin{figure}
\includegraphics{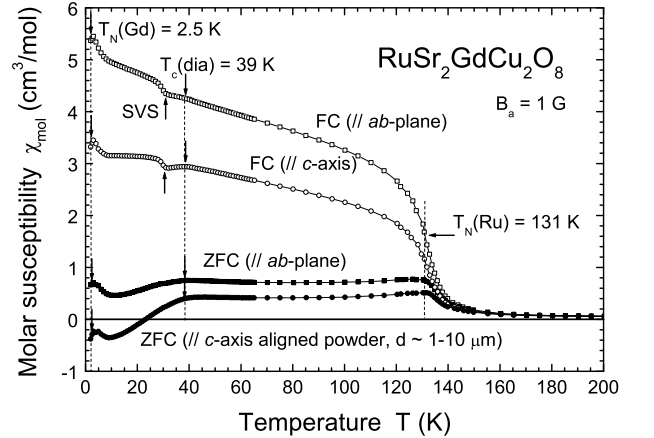}
\caption{\label{label} Low temperature, low field (1-G field-cooled
(FC) and zero-field-cooled (ZFC)) anisotropic magnetic and
superconducting properties of RuSr$_{2}$GdCu$_{2}$O$_{8}$ aligned
powder.}
\end{figure}

\begin{figure}
\includegraphics{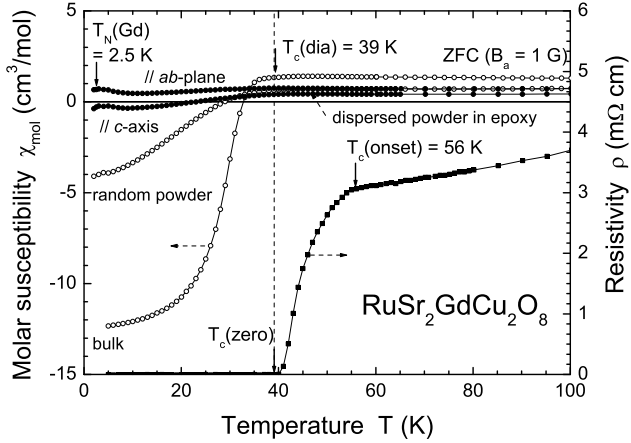}
\caption{\label{label} Low temperature superconducting properties of
aligned powder (dispersed microcrystallines in epoxy), random powder
and bulk RuSr$_{2}$GdCu$_{2}$O$_{8}$ sample.}
\end{figure}

Low temperature, low field (1-G field-cooled (FC) and
zero-field-cooled (ZFC)) anisotropic magnetic and superconducting
properties of RuSr$_{2}$GdCu$_{2}$O$_{8}$ aligned powder are shown
in figure 10. A clear weak-ferromagnetic ordering T$_{N}$(Ru) at 131
K was observed in all data with $\chi_{ab} > \chi_{c}$ in the
weak-ferromagnetic state for both FC and ZFC measurements. A
superconducting diamagnetism setting in at the vortex melting
temperature T$_c$(dia) of 39 K, same as the T$_c$(dia) in the bulk
sample, can be clearly seen in the ZFC measurements but with a
weaker diamagnetic signal. In the FC measurements, the onset of
diamagnetic signal can be recognized as a kink at T$_c$(dia), the
abrupt increase of magnetic signal at 30 K was attributed to the
magnetic field profile inside the SQUID magnetometer \cite{p8} and
corresponds to the spontaneous vortex state temperature T$_{SVS}$,
the same as the one observed in bulk samples.\cite{p6} The
antiferromagnetic Gd$^{3+}$ order is observed at T$_{N}$(Gd) = 2.5
K.

The electrical resistivity data of bulk sample in figure 11
indicates a high superconducting onset temperature of 56 K, with a
much lower T$_c$(zero) = T$_{c}$(dia) at the vortex melting
temperature of 39 K. Slightly larger diamagnetic signal for random
powder is probably due to partially intergrain supercurrent
shielding through partial grain contact. The weak diamagnetic signal
observed in aligned powder-in-epoxy samples below T$_{c}$(dia) is
due to pure intragain shielding with, in addition, long penetration
depth $\lambda$ ($\lambda_{ab} \sim$ 0.55 $\mu$m, $\lambda_{c} \sim$
0.66 $\mu$m) in comparison with powder grain size ($\sim$ 1-10
$\mu$m) and the two-dimensional (2D) character of CuO$_2$ layers.

The anisotropic high-field ($\pm$7 T) isothermal superconducting
hysteresis loops M-B$_{a}$ at 100 K (Fig. 12) indicate the initial
M$_{ab} >$ M$_{c}$ as expected from the susceptibility. Since the
magnetization curves for both field orientations showed
weak-ferromagnetic behavior, the weak-ferromagnetic ordered m(Ru)
dominates over the paramagnetic m(Gd) in the magnetic response and
should be responsible for the complex metamagnetic-like behavior
around 1 T for field applied along the $\it{ab}$-plane.

\begin{figure}
\includegraphics{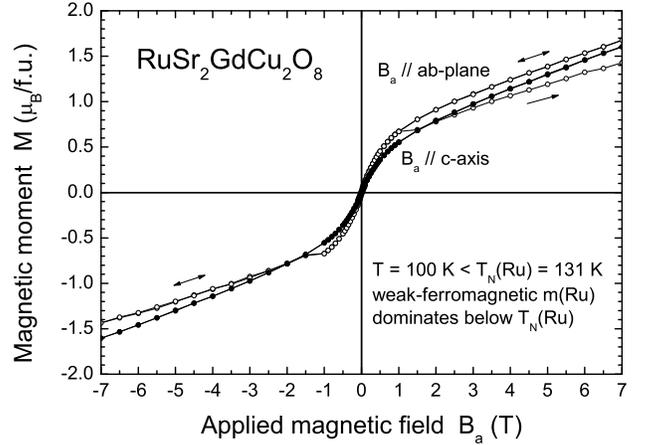}
\caption{\label{label}  Anisotropic high-field hysteresis loop
m$_{c}$(B$_{a}$) and m$_{ab}$(B$_{a}$) at 100 K for aligned
RuSr$_{2}$GdCu$_{2}$O$_{8}$ sample.}
\end{figure}

\begin{figure}
\includegraphics{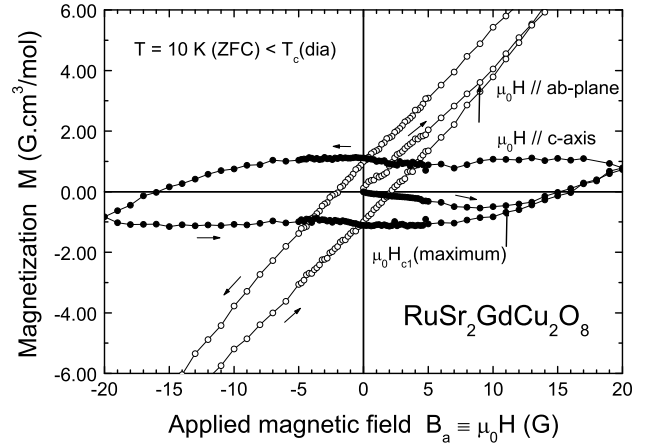}
\caption{\label{label}  Anisotropic high-field hysteresis loop
M$_{c}$(B$_{a}$) and M$_{ab}$(B$_{a}$) at 10 K for aligned
RuSr$_{2}$GdCu$_{2}$O$_{8}$ sample.}
\end{figure}

The magnetization curves with applied field B$_{a}$ along
$\it{c}$-axis, M$_{c}$(B$_{a}$,T), and $\it{ab}$-plane,
M$_{ab}$(B$_{a}$,T), for aligned RuSr$_{2}$GdCu$_{2}$O$_{8}$ powder
at 10 K are shown in Fig. 13. A strongly anisotropic magnetization
was observed for different field orientations. A superconducting
hysteresis loop with a weak paramagnetic background was observed in
B$_a$ $\parallel$ $\it{c}$-axis with a maximum diamagnetic signal at
magnetization peak field $\mu_{0}$H$^{c}$(peak) of 11 G. However, no
superconducting diamagnetic signal was directly observed for the
initial magnetization curve for field applied along $\it{ab}$-plane
due to the much stronger paramagnetic-like background. After
subtracting the paramagnetic-like background, the maximum
diamagnetic signal for M$_{ab}$ can be obtained as
$\mu_{0}$H$^{ab}$(peak) = 9 G. The great difference of magnetization
anisotropy between 100 K and 10 K observed cannot be explained by
simple superposition of magnetic and superconducting components that
suggests complicated interplay between the doped electrons in
weak-ferromagnetically ordered RuO$_6$ layers and superconducting
holes in CuO$_2$-layers.

\section{Conclusion}
Anisotropic powder alignment is achieved for
RuSr$_{2}$RCu$_{2}$O$_{8}$ weak-ferromagnetic superconductors (R =
Eu, Gd, and Gd$_{0.5}$Dy$_{0.5}$). Due to spin-orbital related
short-range anisotropic exchange interaction, paramagnetic
susceptibility $\chi_{ab}$(Ru/Cu) $> \chi_{c}$(Ru/Cu) at 300 K in
RuSr$_{2}$GdCu$_{2}$O$_{8}$ and RuSr$_{2}$EuCu$_{2}$O$_{8}$,
$\it{c}$-axis aligned powder can be achieved only using
field-rotational method. Total $\chi_{ab} < \chi_{c}$ at room
temperature is dominated by R$^{3+}$. Due to long-range Ru
anisotropic exchange interaction, total $\chi_{ab} > \chi_{c}$ were
observed below crossover temperature $\sim$ 185 K, with
weak-ferromagnetic T$_{N}$(Ru) = 131 K, superconducting T$_{c}$(dia)
= 39 K and T$_{N}$(Gd) = 2.5 K. Weak diamagnetic signal observed
below T$_{c}$(dia) was due to pure intragain shielding with long
penetration depth $\lambda$ ($\lambda_{ab} \sim$ 0.55 $\mu$m,
$\lambda_{c} \sim$ 0.66 $\mu$m) and the two-dimensional (2D)
character of CuO$_2$ layers.

\begin{acknowledgments}
This work was supported by the National Science Council of R.O.C.
under contract Nos. NSC95-2112-M-007-056-MY3, NSC95-2112-M-032-002,
and NSC97-2112-M-003-001-MY3.
\end{acknowledgments}


\end{document}